# Chemical Doping and Enhanced Solar Energy Conversion of Graphene/Silicon Junctions

Xinming LI, Hongwei ZHU*, Kunlin WANG, Jinquan WEI, Guifeng FAN, Xiao LI, Dehai WU

*Key Laboratory for Advanced Manufacturing by Material Processing Technology, Ministry of Education*
*Department of Mechanical Engineering, Tsinghua University Beijing 100084, P. R. China*
Email: *hongweizhu@tsinghua.edu.cn

**Abstract:** The effect of chemical doping of graphene films on the photovoltaic properties of the graphene/silicon Schottky junction solar cells was investigated. Thionyl chloride modification greatly enhanced the conductivity of graphene film, resulting in a significant improvement in cell performance with a 3-fold increase in conversion efficiency (up to 3.9%) and good short-term stability.

## 1. Introduction

Graphene, a novel two-dimensional (2D) carbon nanomaterial, has attracted a widely attention since it discovery in 2004 [1]. In the past few years, tremendous progress has been achieved in graphene synthesis, characterizations and applications. For example, graphene has been produced in the form of ultra-thin sheets consisting of one or a few atomic layers by chemical vapor deposition (CVD) [2-5] or solution processing [6-9], and can be transferred to various substrates. Graphene has been widely used for composites, nanoelectronics and transparent electrodes owing to its unique 2D nanostructure and conductibility [10].

One immediate use of graphene is to make flexible, transparent, conductive electrodes by taking advantage of a combination of its extremely high optical transparency and electrical conductivity. For example, solution-processed graphene dispersed into polymers such as poly(3-octyl thiophene) (P3OT) or poly (3-hexylthiophene) (P3HT) was used as acceptor materials [11-13]. Graphene films were also used as conductive and transparent electrodes to replace ITO in organic [14-16] and dye-sensitized solar cells [17,18]. CVD synthesized graphene showed higher conductivity than solution-processed graphene. However, the energy conversion performance of these graphene-based cells is still low because of the quality limitation of graphene. The presence of surface wrinkles, defects has severely suppressed its condutivity, in turn its practical use as electrode materials.

Recently, we proposed a photovoltaic model in which highly conductive, semi-transparent graphene films is coated on n-type silicon (n-Si) wafer to form Schottky junction [19]. Our results showed that in this Schottky solar cell, graphene as energy conversion materials not only contributes to charge separation and transport, but also functions as transparent electrode.

As mentioned above, the conductibility of graphene films remains one of the most important restriction factors because of its discontinuity and the presence of surface wrinkles. The power conversion efficiencies of the graphene/n-Si solar cells are still low (~1.6% at AM 1.5). Though the partially replacement devices are still far lower in efficiency than pure silicon cells, this simple concept of Schottky junctions made of graphene and n-Si, with a good understanding of surface passivation, doping, and junction formation, will lead to much more efficient and stable graphene-based solar cells.

Previous work has shown that chemical doping of carbon nanotubes (CNTs) or graphene could result in the remarkable increase of it conductivity, mainly promoting the charge transport [20-24]. Kong et al. demonstrated that the work function of graphene can be tuned by $AuCl_3$ doping [22]. Kaner et al. used thionyl chloride ($SOCl_2$) vapor treatment for a hybrid layer of CNTs and chemically converted graphene films, and greatly increased the film conductivity [23]. Chhowalla et al. also reported that $SOCl_2$ treatment could reduce the sheet resistance of graphene oxide films, making the transparent thin films as the hole collecting electrodes in organic photovoltaic devices [24]. These results suggested that $SOCl_2$ chemical doping of graphene films is possible for improve the power conversion efficiencies.

In light of these studies, Schottky solar cells make from $SOCl_2$ doped graphene and n-Si were tested, showing a 3-fold enhancement in efficiency (up to 3.9%) with a good short-term stability.

## 2. Experimental

Assembly of the graphene/silicon (G/Si) Schottky solar cells involves three steps, as shown in Figure 1. First, graphene films were synthesized by CVD method using nickel foils as substrates. To obtain continuous films, nickel substrate was etched away with a $FeCl_3$ aqueous solution. Second, the freestanding graphene films were transferred onto n-Si substrates with pre-patterned electrodes to make Schottky junction solar cells. Third, the cells were exposed in the SOCl$_2$ vapor for a few seconds and acyl chloride groups were doped on the films. The photovoltaic properties of the solar devices were measured under AM 1.5 illumination using a Newport solar simulator.

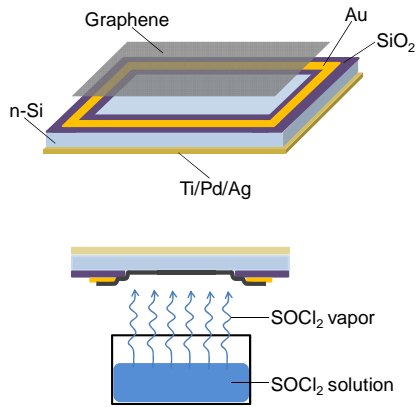

**Figure 1. Schematic diagrams of the G/Si solar cell and SOCl$_2$ dopping.**

## 3. Results and discussion

Scanning electron microscopy (SEM) and transmission electron microscopy (TEM) characterizations show the surface morphology of the graphene films before and after SOCl$_2$ treatment (Figure 2). The pristine graphene films are clean and have many wrinkles on the surface owing to the nucleation of defect lines on Ni terrace edges and thermal-stress-induced step edges and defect lines. The organic substance is observed after the graphene films were exposed in the SOCl$_2$ vapor. Upon SOCl$_2$ treatment, fairly good adhesion of functional groups to films was obtained and some wrinkles were eliminated because of the favorable chemical doping. It is also found that upon SOCl$_2$ doping, the sheet resistance of graphene films is remarkably reduced to 20% of its original value after only 20s treatment.

Figure 3a shows the current density-voltage ($J$-$V$) curves before and after SOCl$_2$ treatment of a typical G/Si solar cell. The open-circuit voltage ($V_{oc}$), short-circuit current density ($J_{sc}$) and fill factor (FF) of the pristine solar cell are 0.42V, 12.7 mA/cm$^2$ and 35 %, corresponding to a conversion efficiency ($\eta$) of 1.84%. Upon SOCl$_2$ treatment, these parameters are significantly increased: $V_{oc}$ to 0.52V, $J_{sc}$ to 13.2 mA/cm$^2$ and FF to 58%. As a result, the $\eta$ of the treated cell reached 3.93%. We examined a few of samples to confirm the effect of the chemical treatment, as listed the Table 1. The photovoltaic performance including $V_{oc}$, $J_{sc}$ and FF after treatment was improved obviously and corresponded to an overall $\eta$ of 3.7~3.9%, which is about 2~3.7 times higher than those for the cells without chemical doping.

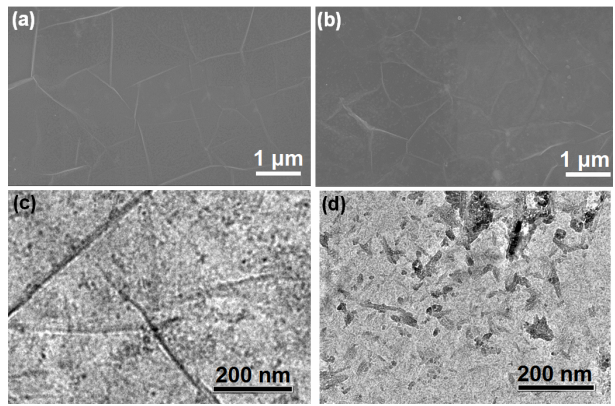

**Figure 2. SEM and TEM characterizations of graphene films before (a, c) and after (b, d) chemical doping.**

The effect of SOCl$_2$ doping time on the G/Si solar cells was investigated. It is found that the $\eta$ was sharply increased and tended to be stable quickly when the film was doped with SOCl$_2$ (Figure 3b). The maximum $\eta$ is obtained at 800s of treatment. However, the $\eta$ begins to gradually reduce after prolonged SOCl$_2$ doping over 800s. In addition, $V_{oc}$, $J_{sc}$ and FF show similar variations. Figure 3c, d show the effect of SOCl$_2$ doping time on $V_{oc}$, $J_{sc}$, FF and $\eta$. Evidently, the $\eta$ of the solar cell is mainly determined by FF. Meanwhile, FF is also increasing with the doping time and falls down with prolonged SOCl$_2$ doping of more than 800s.

The high conversion efficiency can be attributed to the enhancement of G/Si Schottky junction, which makes charge separation and transport happen more easily. First, the graphene films were contacted to the silicon favorably owing to the SOCl$_2$ doping. Second, chemical doping of SOCl$_2$ significantly enhances the conductivity of the graphene films.

The non-linear $J$-$V$ characteristic of the Schottky junction can be expressed as [25]:

$$I = I_s [\exp\left(\frac{eV \text{-} eIRs}{nkT}\right) - 1]$$

where $J_s$ is the reverse saturation current density, $e$ is electronic charge, $n$ is the diode ideality factor, $k$ is the Boltzman constant, $T$ is the absolute temperature, $R_s$ is the series resistance of the solar device. The FF is greatly dependent on $R_s$ and $n$, as shown in Figure 5. At the initial stage of the SOCl$_2$ doping, $R_s$ and $n$ were reduced obviously, leading to an increase in FF. After 800s treatment, $R_s$ and $n$ reached 8.4Ω and 1.54, respectively. However the overlong chemical treatment would induce the oxidation of silicon, which in turn degraded the ideal factor and reduce the conversion efficiency.

Finally, the short-term stability of the SOCl$_2$ doped G/Si cell was investigated. Degradation of the cell performance was observed with the gradual dropdown of $V_{oc}$, $J_{sc}$, FF and $\eta$. After one week, the cell efficiency was stabilized at ~2.5%, which is still higher than the original value before doping. This decrease is probably because of the volatility of SOCl$_2$, resulting in a decrease in FF.

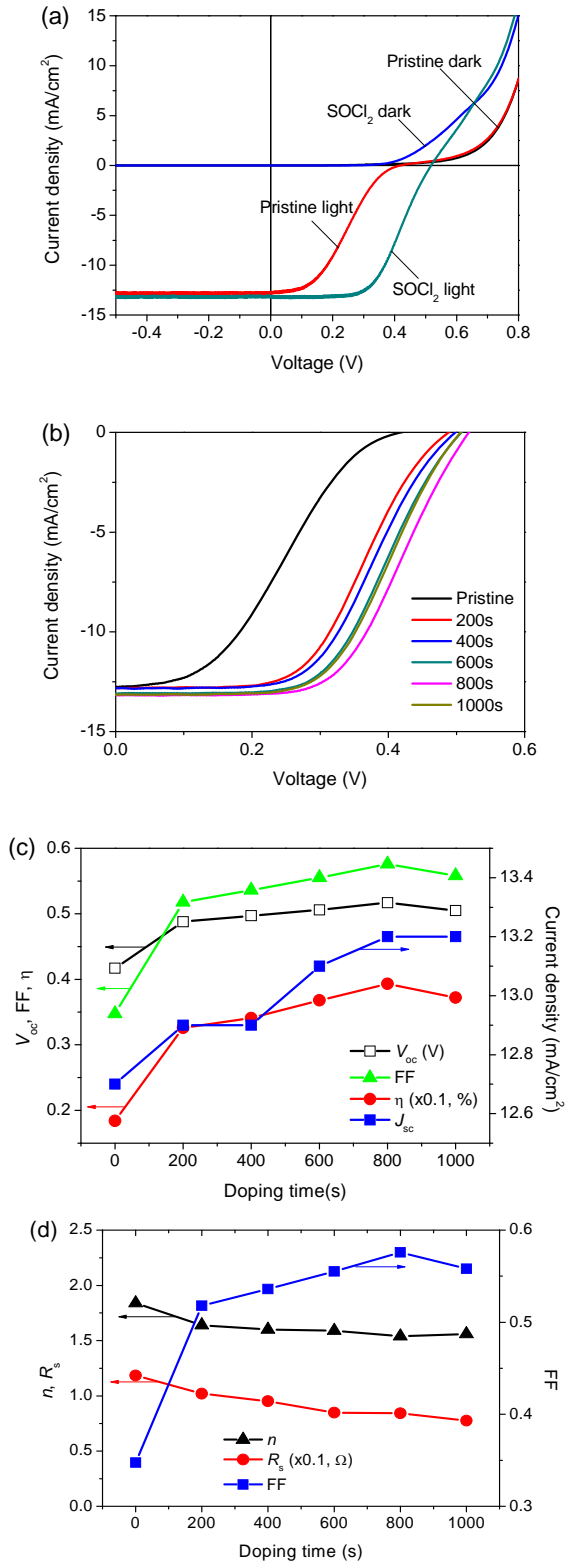

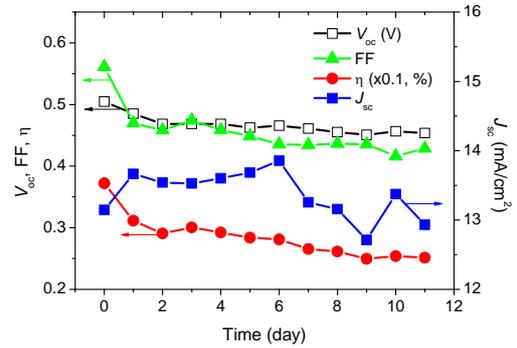

**Figure 3.** Photovoltaic characterizations of doped G/Si solar cells. (a) Dark and light *J-V* curves of the cells illuminated at AM 1.5 before and after doping. (b) Light *J-V* curves of the cell of different SOCl$_2$ doping time. (c) $J_{sc}$, $V_{oc}$, FF and η plots as a function of doping time. (d) $R_s$ and *n* as a function of doping time.

**Figure 4.** Short-term stability ($J_{sc}$, $V_{oc}$, FF and *η* variations) of a doped G/Si solar cell.

**Table 1.** Photovoltaic performance of two typical cells before and after SOCl$_2$ doping.

| Samples | Photovoltaic performance | | | |
|---|---|---|---|---|
| | $V_{oc}$ (V) | $J_{sc}$ (mA/cm$^2$) | FF | η (%) |
| G/Si01, pristine | 0.433 | 11.63 | 0.20 | 1.01 |
| G/Si01, doped | 0.520 | 13.4 | 0.54 | 3.75 |
| G/Si02, pristine | 0.417 | 12.7 | 0.35 | 1.84 |
| G/Si02, doped | 0.517 | 13.2 | 0.58 | 3.93 |

## 4. Conclusions

We studied the effect of chemical doping of graphene films on the photovoltaic performance of the G/Si Schottky junction solar cells. Our results show that the chemical doping of graphene films with SOCl$_2$, led to an obvious enhancement of Schottky junction, which is attributed to the drop of the sheet resistance of graphene film. The solar energy conversion efficiency was improved to 3.7~3.9%, which are about 2~3.7 times higher than those for the cells without chemical doping. Further improvement of G/Si solar cells may be envisioned via further optimization of the graphene films and chemical treatments.

## 5. Acknowledgement

This work was supported by National Science Foundation of China (No. 50972067), Tsinghua National Laboratory for Information Science and Technology (TNList) Cross-discipline Foundation and Program for New Cen-